\documentclass[12pt]{article}
\usepackage{epsfig}
\usepackage{cite}
\def\beq{\begin{equation}}
\def\eeq{\end{equation}}
\def\barr{\begin{array}}
\def\earr{\end{array}}
\def\dis{\displaystyle}

\def\rw{{\cal R}_w}

\def\vbar{\rho}

\newcommand{\lsim}
{{\;\raise0.3ex\hbox{$<$\kern-0.75em\raise-1.1ex\hbox{$\sim$}}\;}}
\def\fcap{{\;\raise1.8ex\hbox{{\tiny $\cap$}\kern-0.7em\raise-1.8ex\hbox{$f$}}}}
\def\fdc{{\;\raise2.0ex\hbox{{\footnotesize $\cap$}\kern-0.8em\raise-2.0ex\hbox{$\fcap$}}}}

\begin{document}
\thispagestyle{empty}

\begin{center}
{\Large\bf Living on the edge in a spacetime with multiple warping}\\[20mm]
Debajyoti Choudhury\footnote{E-mail: debchou@physics.du.ac.in}\\
{\em Department of Physics and Astrophysics\\University of Delhi\\
Delhi 110 007, India} \\[8mm]

Soumitra SenGupta\footnote{E-mail: tpssg@iacs.res.in} \\
{\em Department of Theoretical Physics\\ Indian Association for the
Cultivation of Science\\
Calcutta - 700 032, India} \\[20mm]
\end{center}

 
\bigskip
\abstract{The Randall-Sundrum warped braneworld model is generalised
to six and higher dimensions such that the warping has a non-trivial
dependence on more than one dimension. This naturally leads to a
brane-box like configuration alongwith scalar fields with possibly
interesting cosmological roles. Also obtained naturally are two towers
of 3 branes with mass scales clustered around either of Planck scale
and TeV scale.  Such a scenario has interesting phenomenological
consequences including an explanation for the observed hierarchy in
the masses of standard model fermions.}

\newpage
\section{Introduction}
Theories with extra spacetime dimensions have drawn considerable
attention since the original proposals of Kaluza and Klein. There has
been a renewed interest in such theories since the emergence of string
theory. Several new ideas in such directions evolved to explore
various implications of the presence of extra spatial dimensions in
the context of particle phenomenology and
cosmology\cite{arkani,antoniadis,witten,randall,lykken,cohen,kaloper}. Although
these ideas are not necessarily derived from string theory, there have
been serious efforts towards establishing links between the two.  Two
of the most prominent such extra dimensional theories developed in the
context of the braneworld models are due to Arkani-Hamed, Dimopolous,
Dvali (ADD)\cite{arkani} on the one hand and Randall and Sundrum
(RS)\cite{randall} on the other.  While the ADD model involves the
presence of large compact extra dimensions (the radius(radii) of the
extra compact dimension(s) being much larger than the Planck length),
the RS model proposes the existence of a warped geometry in $3+1$
dimensions in the background of a $4+1$ dimensional anti de-Sitter
(AdS) bulk.  Both the theories claim to solve the so called
``naturalness problem'' in standard model, which originates from the
need of an unnatural fine-tuning of parameters of the theory to
stabilize the Higgs mass within the TeV scale against large radiative
corrections.  Although the presence of large radii in the ADD model
indicates the reappearance of the hierarchy in a different guise, the
RS scenario is apparently free from such problems.  However, in this
case the braneworld model itself is not stable and it was first shown
by Goldberger and Wise (GW)\cite{gw} that by introducing a scalar
field in the bulk, the modulus---namely, the brane separation---in RS model
can be stabilized without the need of any unnatural
fine-tuning. Assumption of a negligibly small scalar back-reaction on
the metric in the GW approach prompted further work in this direction
where the modifications of the RS metric due to back-reaction of bulk
fields have been derived\cite{csaki}. The stability issues in such
cases have been re-examined as also the effects of other bulk fields
like gauge field or higher form fields been studied in several
works~\cite{ssg,ferreira,bulk,cvetic}.  Such warped geometries are
expected to have additional consequences in particle phenomenology
over and above the hierarchy issue.

As a natural extension to the RS scenario with one extra spatial dimension,
several extensions of the RS model to more than one extra dimension
have been proposed~\cite{rshigh}. In particular some cosmological implications
of warped geometry in six dimension have been explored in the context of
dynamical compactification of extra dimension \cite{papantonopoulos}. 
Most of these consider the presence of 
several independent $S_1/Z_2$ orbifolded dimensions along with $M_4$. 
We, however,
propose a more intricate scenario wherein the warped compact dimensions get
further warped by a series of successive warping 
leading to  multiple warping of the space-time with various p-branes 
sitting at the different orbifold fixed points satisfying appropriate
boundary conditions. Various lower dimensional branes along with the standard model 3-brane 
exist at the intersection edges of the higher dimensional branes.    
Thus the resulting geometry of the $D$-dimensional space-time is 
$M^{1,D-1} \rightarrow \left\{ [M^{1,3} \times S^1/Z_2] \times 
   S^1/Z_2 \right\} \times \cdots$, with
$(D - 4)$ such warped directions.
A series of scales are thus generated from each of these successive 
warpings and 
we show that such a spacetime with multiple warping leads to
interesting phenomenology.

The original RS model corresponded to a 
5-dimensional AdS spacetime wherein the extra dimension was $S^1/Z_2$
orbifolded. Two branes (called the standard model brane and Planck
scale brane) were placed at the two orbifold fixed points and
appropriate brane tensions at the boundaries of the orbifold were
determined in terms of the bulk cosmological constant.  In such a
model, it was shown that a TeV scale can be generated at the standard
model brane without any unnatural fine tuning of parameters.  We
propose here a model in a ($5+1$)-dimensional bulk AdS spacetime where
both the extra coordinates are compactified in succession on circles with
$Z_2$ orbifoldings. We show that the six dimensional Einstein's
equation can be solved exactly for such a geometry and that the
resulting solution for the metric is doubly warped.  Although the
warping of the metric along one of the compact coordinates resembles
exponential warping as found by Randall and Sundrum, the other one
turns out to be a hyperbolic warping.  Such a solution with doubly
orbifolded boundary conditions results in a box-like picture of the
bulk, where the walls of the box are ($4+1$)-dimensional branes.  The
$Z_2$ orbifoldings along the two compact coordinates puts stringent
conditions on the brane tensions.  Four ($3+1$)-dimensional branes are
formed at the four edges of the intersecting 4-branes.  We can then
identify our $3+1$ dimensional standard model brane with one of the
edges by requiring the desired TeV scale while the Planck scale brane
resides at another edge. We thus live at one edge of the proposed
spacetime ith multiple warping. The other two edges correspond to two more
$3+1$ dimensional branes with some intermediate energy scales.  With 
the aforementioned $Z_2$ orbifoldings further requiring
coordinate-dependent brane tensions,  a possible origin for the same
is proposed.  As could be expected, the mass
of a standard model scalar field like the Higgs boson 
is doubly warped from the
Planck scale down to the TeV scale and thereby the fine tuning problem is
resolved. The crucial aspect of the double warping scenario turns out
to be unequal warping in two directions. We show that while warping in
one direction is large, the other is necessarily small.

We then generalise our result to a large number of extra dimensions,
where several scales are generated and because of unequal warping,
half of these are clustered around Planck scale while the rest are
around TeV scale. We then argue that such a clustered scale brane
model can offer a possible explanation of the observed mass
differences in the standard model fermions.

\section{Six dimensional doubly warped spacetime}
     \label{sec:6d}

The spacetime that we are interested in is a doubly compactified
six-dimensional one with a $Z_2$ orbifolding in each of the compact
directions. In other words, the manifold under consideration is
$M^{1,5} \rightarrow [M^{1,3} \times S^1/Z_2] \times S^1/Z_2$.  To set
the notation, the non-compact directions would be denoted by $x^\mu \,
(\mu = 0..3)$ and the orbifolded compact directions by the angular
coordinates $y$ and $z$ respectively with $R_y$ and $r_z$ as
respective moduli. The corresponding metric is,
\beq
ds^2 = b^2(z)[a^2(y)\eta_{\mu\nu}dx^{\mu}dx^{\nu} + R^2_y dy^2] + r^2_z dz^2
    \label{metric} 
\eeq 
where $\eta_{\mu\nu} = diag(-1,1,1,1)$. Since orbifolding, in general,
requires a localized concentration of energy, we introduce four
4-branes ($4+1$ dimensional objects) at the orbifold fixed ``points'',
namely $y = 0, \pi$ and $z = 0, \pi$.

The total bulk-brane action is thus given by,
\beq
\barr{rcl}
S & = & \dis S_6 + S_5 + S_4 \\[1ex]
S_6 & = & \dis \int {d^4 x} \, {d y} \, {d z} \, 
          \sqrt{-g_6} \; \left(R_6 - \Lambda \right) 
\\[2ex]
S_5 & = & \dis \int {d^4 x} \, {d y} \, {d z} \, 
           \left[ V_1 \, \delta(y) + V_2 \, \delta( y - \pi) \right]
\\[1.5ex]
    & + & \dis \int {d^4 x} \, {d y} \, {d z} \, 
           \left[ V_3 \, \delta(z) + V_4 \, \delta(z - \pi) \right]
\\[2ex]
S_4 & = & \dis \int d^4 xdydz \sqrt{-g_{vis}}[{\cal L} - \hat V] \ .
\earr
    \label{Action}
\eeq
Note that, in general, we have, for the brane potential terms $V_{1,
2} = V_{1, 2}(z) $ whereas $V_{3, 4} = V_{3, 4}(y) $. The presence of
the term $S_4$ indicates the contributions due to possible 3-branes
located at $(y, z) = (0,0), (0, \pi), (\pi, 0), (\pi, \pi)$.

The full six dimensional Einstein's equation can be written as,
\beq
\barr{rcl}
- \dis M^4 \, \sqrt{-g_6} \, \left( R_{MN} - \, \frac{R}{2} \, g_{MN}\right)
& = & \dis \Lambda_{6} \, \sqrt{-g_6} \, g_{MN} \\
& + & \dis 
\sqrt{-g_5} \, V_{1}(z) \, g_{\alpha\beta} \, \delta^{\alpha}_{M} \, 
                 \delta^{\beta}_{N} \, \delta(y)  
 +  \dis \sqrt{-g_5} \, V_{2}(z) \, g_{\alpha\beta} \, \delta^{\alpha}_{M}
                  \, \delta^{\beta}_{N} \, \delta(y-\pi) 
 \\[1ex]
&+ & 
  \sqrt{-\widetilde g_5} \, V_{3}(y) \, \widetilde g_{\tilde\alpha \tilde\beta} \,
\delta^{\tilde\alpha}_{M} \, \delta^{\tilde\beta}_{N} \, \delta(z) 
 + \sqrt{-\widetilde g_5} \, V_{4}(y) \, 
      \widetilde g_{\tilde\alpha \tilde\beta} \, \delta^{\tilde\alpha}_{M}
      \, \delta^{\tilde\beta}_{N} \, \delta(z-\pi)
\earr
\eeq
Here M,N are bulk indices, $\alpha,\beta$ run over the usual four
spacetime coordinates ($x^\mu$) and the compact coordinate $z$ while
$\tilde{\alpha},\tilde{\beta}$ run over $x^\mu$ and the compact
coordinate $y$.  And, finally, $g,\widetilde{g}$ are the respective
metrics in these (4+1)-dimensional spaces.

On substituting our ansatz for the metric (eqn.\ref{metric}),  
the $yy$ and $zz$ components of Einstein's equations reduce to a set of two simpler equations, namely
\beq
\barr{rcl}
\dis
 2 \, M^4 \, \left[3 \, r^2_z \, a'^2 + 3 \, R^2_y \, a^2 \, {\dot b}^2 +
  2 \, R^2_y \, a^2 \, b \, \ddot{b} \right] & = & -b^2 \, a^2 \, r_z \, R^2_y 
   \left[ r_z \, \Lambda +  V_3 \, \delta(z) + V_4 \, \delta(z-\pi) \right]
\\[1.5ex]
2 \, M^4 \, \left[3 \, r^2_z \, a'^2 + 5 \, R^2_y \, a^2 \, {\dot b}^2 +
  2 \, r^2_z \, a \, a''\right] & = & -b \, a^2 \, r^2_z \, R_y 
   \left[ R_y \, \Lambda \, b + 
 V_1 \, \delta(y) + V_2 \, \delta(y-\pi) \right] 
\earr
    \label{Einstein:comp}
\eeq
where primes denote differentiation w.r.t. $y$, while 
dots denote differentiation w.r.t. $z$. Starting with 
the bulk part of eqn.(\ref{Einstein:comp}), and 
rearranging terms, we have
\beq
\frac{a'^2}{a^2} = c^2 = R_y^2 \, \left[
\frac{\dot{b}^2}{r_z^2} + \frac{2 \, b \ddot{b}}{3 \, r_z^2}
 + \frac{b^2 \, \Lambda}{6 \, M^4} \right]
\eeq
where $c$ is an arbitrary constant. The solution to the above is given by
\beq
\barr{rcl c rcl }
a(y) & = & \exp(-c \, y) & \qquad & 
b(z) &=& \dis \frac{\cosh(k \, z)}{\cosh(k \, \pi)}
\\[2ex]
c & \equiv & \dis \frac{R_y \, k}{ r_z \, \cosh(k \, \pi)}
& &  k & \equiv &  \dis r_z \, \sqrt{\frac{-\Lambda}{10 \, M^4}}.
\earr
\label{soln:interm}
\eeq
It can be easily ascertained that eqn.(\ref{soln:interm}) satisfies
each of the two eqns.(\ref{Einstein:comp}) as long as they are
restricted to the bulk. As is quite apparent from the form of the
solution, the presence of an exponential warping (as in RS model) in
the $y$- direction necessitates a negative value for the bulk
cosmological constant $\Lambda$, thereby signalling an AdS bulk. Of
course, for $c^2 < 0$, an alternative (oscillatory) solution for both
$a(y)$ and $b(z)$ are possible. However, this, manifestly, does not
lead to the desired warping of the spacetime metric and hence shall
not be considered any further. Similarly, we discount solutions of the
form $b^2(z) = - (c^2 \, r_z^2 / k^2 \, R_y^2) \, \sinh^{2}(k \, z)$ as
this leads to a bulk metric with a $(4, 2)$ signature.

Note that the $Z_2$ orbifolding in the $y$-direction, namely $y \equiv
- y$, demands that $a(y) = \exp(- c \, |y|)$ whereas the symmetric form
of $b(z)$ obviates the need for an analogous requirement.  The full
metric thus takes the form
\begin{eqnarray}
ds^2 & = & \dis \frac{\cosh^2(k \, z)}{\cosh^2 (k \, \pi)} \,
  \left[ \exp\left(- 2 \, c \, |y| \right)
                \, \eta_{\mu \nu} \, d x^\mu \, d x^\nu 
     + R_y^2 \, d y^2 \right] + r_z^2 \, d z^2 \ .
\end{eqnarray}

Next, we focus our attention on the boundary terms so as 
to determine the brane tensions.\\  
Using $a(y) = exp(-c|y|)$ , substituting for $c$ from equ.
(\ref{soln:interm}) and integrating the second of eqns.(\ref{Einstein:comp}) over an infinitesimal 
interval across the two boundaries at $ y = 0 ,y= \pi $ respectively, we
obtain
\beq
V_1(z) = - V_2(z) = 8M^2 \sqrt{\frac{-\Lambda}{10} } \, 
{\rm sech}(k \, z) \ .
\eeq
In other words, the two 4-branes sitting at $y = 0$ and $y = \pi$ have 
$z$-dependent tensions, a feature that will return to in the next section. The
fact that these tensions are equal and opposite is reminiscent of the 
original RS-form and is but a consequence of the exponential warping 
and the fact of these branes sitting at the orbifold fixed ``points''.

Similarly, starting with the first of eqns.(\ref{Einstein:comp}),
using the solution for $b(z)$ from (\ref{soln:interm}) 
and integrating over an infinitesimal interval across $ z = 0 $, we find,
\beq
V_3(y) = 0
\eeq
This, of course, was to be expected since the 
smooth behaviour of $b(z)$ as $z \to 0$ obviates the necessity 
for any localized energy density at $z = 0$. 
On the other hand, integrating over an infinitesimal interval 
across $z = \pi$ gives 
\beq
V_4(y) = -\frac{8 \, M^4 \, k}{r_z} \, \tanh(k\pi)  \ ,
\eeq
a constant, unlike the case for $V_{1,2}(z)$, but quite similar to the case
for the original RS model. This, again, is not unexpected, for $V_{3,4}$ 
were introduced to account for the orbifolding in the $z$-direction and 
with $g_{zz}$ being a constant, the corresponding hypersurfaces should 
only have a constant energy density. The fact of $g_{yy}$ being a non-trivial 
function of $y$, however, made it mandatory that the two hypersurfaces 
accounting for the $y$-orbifolding must have a $z$-dependent energy 
density.

\subsection{Brane identities and the extent of warping}
      \label{brane-id}
We have thus determined the tensions for all of the 4-branes in the theory. 
As indicated earlier, the intersection of two 4-branes may be identified 
with a 3-brane with a tension that, to the leading order,
 is an algebraic sum of the energy 
densities  contributed by each of the 4-branes. With this 
identification, the theory, thus, contains four 3-branes located at
$(y, z) = (0,0), (0, \pi), (\pi, 0), (\pi, \pi)$.

With the 3-brane located at $(y = 0, z = \pi)$ suffering no warping of
the metric on it, it is logical that it be identified with the Planck
brane. Note that there is no unique assignement of the Standard
Model(visible) brane! Each of the other three offers a valid choice
depending on the values of the parameters ($k, c$). The latter, of
course, are determined in terms of $\Lambda, r_z$ and $R_y$, with the
6-dimensional Planck mass $M$ being essentially the same as the
4-dimensional one because of the relation,
\beq
M_{P}^{2} \sim \frac{M^{4} \, r_z \, R_y}{2 \, c \, k}\,
           \left [1- e^{-2c\pi}\right]\, \left[\frac{tanh(k\pi)}{cosh^2(k\pi)} + \frac{tanh^3(k\pi)}{3}\right]
\eeq
Each choice would have its own unique
phenomenological consequences. If we adopt the conservative view that
there exists no other brane with a natural energy scale lower than
ours, we must identify the SM brane with the one at $y= \pi, z = 0$.
For such a choice, 
\beq
V_{\rm vis} = -8M^2\sqrt{\frac{-\Lambda}{10}} \, \qquad 
V_{\rm Planck} = 8M^2\sqrt{\frac{-\Lambda}{10}} \; [{\rm sech}(k \, \pi) - { \rm tanh}{k \, \pi}] \, 
\label{brane_tensions}
\eeq
with the two other 3-branes located at $(0,0)$ and $(\pi, \pi)$ having
tensions intermediate to the above. Note that whereas the Planck-brane 
must always have a positive tension (given by eqn.\ref{brane_tensions}), 
it is not mandatory that the SM brane must be a negative tension one. For 
example, we could have  identified the latter with the one at 
$(0, 0)$ with the consequence that now, $V_{\rm vis} \simeq
 V_{\rm Planck}$ but 
by paying the price of having at least one brane---that corresponding 
to our present choice---having a lower energy scale.

Before ending this section, we examine 
the possible mass warping in the scalar sector of
the standard model, or, in other words, 
the status of the naturalness problem in such a scenario. 
With the action for  a free scalar propagating on the visible brane being 
given by
\beq
S_{H} = \int d^4x \, \sqrt{-g_{vis}} \;
   \left[ g^{\mu\nu}_{vis} \, D_{\mu}H \, D_{\nu}H - m_0^2 \,
H^2\right] \ ,
\eeq 
a Planck scale mass $m_0$ is warped to 
\beq m = m_0 \,
\frac{r_z \, c}{R_y \, k } \, \exp(-\pi \, c) 
          = m_0 \,  \frac{\exp(-\pi \, c)}{\cosh (k \, \pi)}
\eeq 
on the TeV brane which is quite akin to (but not exactly the same as)
the RS case.  An important point to note is that if we want a
substantial warping in the $z$-direction (from $z = 0$ to $z = \pi$),
$k \, \pi$ must be substantial, i.e. of same order of magnitude as in
the usual RS case. But with $c$ being determined by
eqn.(\ref{soln:interm}), this immediately means that $c$ must be small,
unless there is a large hierarchy between the moduli $r_z$ and $R_y$.
This, in turn, means that we cannot have a large warping in
$y$-direction as well without introducing a new and undesirable
hierarchy.  Similarly, if we demand a large hierarchy in the
$y$-direction (a situation very close in spirit with RS), we must
necessarily live with a relatively small $k \lsim {\cal O}(1)$ and
hence little warping in the $z$-direction. 

An interesting consequence that emerges from this is that, of the two
branes located at ($y= 0, z =0$) and ($y = \pi , z = \pi$), one 
must have a natural mass scale close to the Planck scale, while for the 
other it is close to the TeV scale.  The latter statement 
immediately points to phenomenologically interesting possibilities which shall
be addressed later.
   
\section{Origin of the coordinate dependent brane tension}

We have noticed in the previous section that two of the (4+1)-dimensional 
brane tensions namely $V_{1,2}$ are functions of the
coordinate $z$. While such coordinate-dependence might seem
counterintuitive at first, it should be realized that the Israel
junction conditions only stipulate that there be a concentration of
energy-density at the $y=0, \pi$ hypersurfaces and that these
distributions must have the stipulated $z$-dependence.  A particularly
simple mechanism for arranging such an energy concentration has its
origin in a scalar field confined to the respective branes.

Consider a scalar $\varphi$ on one 4-brane, say on
the brane at $y = y_0$, (where $y_0$ is either $0$ or $\pi$)
with a potential $\cal{V}(\varphi)$. Since the metric on this brane is
\beq 
\barr{rcl} 
d s^2 & = & \dis b_0^2 \, \cosh^2(k \, z) \, \eta_{\mu
\nu} d x^\mu d x^\nu + r_z^2 \, d z^2 \\[2ex] 
b_0 & \equiv & \dis
e^{-2 \, c \, |y_0|} \; \cosh^{-2} (k \, \pi) \ ,
\earr
    \label{soln_on_brane}
\eeq
the action for the scalar is
\beq
S_\varphi = \int d^4 x \, d z \; \sqrt{-g_5} \, 
    \left[g_5^{A B} \, \partial_A \varphi \, \partial_B \varphi + {\cal V}(\varphi) \right] \, 
\eeq
with $g_5^{A B}$ being given by eqn.(\ref{soln_on_brane}). 
This leads to an equation of motion of the form
\beq
r_z^2 \,\frac{\partial {\cal V}}{\partial \varphi}   
= 8 \, k \, \tanh (k z) \, \varphi' 
+ 2  \,  \varphi''
\eeq
where the primes denote differentiation with respect to $z$.
\vspace*{2ex}
Denoting
\beq
{\cal V}(\varphi(z)) \equiv \vbar(z)  \qquad \Longrightarrow  \qquad
\frac{\partial {\cal V}}{\partial \varphi}   = \frac{\vbar'}{\varphi'} \ ,
    \label{def_rho}
\eeq
the equation of motion becomes
\beq \dis 
r_z^2 \,\vbar' \; \cosh^8(k z) 
   = \frac{d}{d z} \, \left[\cosh^8(k z) \; (\varphi')^2 \right] \ .
   \label{eom_phi}
\eeq
Since, for the energy density to give the required brane tension, we must have
\beq
\rho(z) + \left(\frac{\varphi'}{r_z}\right)^2 = V_{1,2} \ ,
    \label{energy_equality}
\eeq
as the case may be,  we now need to find simultaneous solutions of  eqns.
(\ref{eom_phi}\&\ref{energy_equality}) for each of $V_{1,2}$. 

Concentrating first on the 4-brane at $y = \pi$ we find,
\beq
\rho(z)
 = v_0 \; \left[ -\frac{7}{6} \, {\rm sech}(k \, z) 
                         + \xi \; {\rm sech}^4(k \, z) \right] \ ,
\qquad v_0 \equiv 8M^2 \sqrt{\frac{-\Lambda}{10}} \ ,
    \label{rho_1}
\eeq
where $\xi$ is a constant of integration,  and 
\beq
\frac{\varphi'^2}{r_z^2} = v_0 \, \left[ \frac{1}{6} \, {\rm sech}(k \, z) 
                               - \xi \; {\rm sech}^4(k \, z) \right] \ .
   \label{varphi_soln}
\eeq
Positivity of the right hand side (over the entire 4-brane) requires
\beq
\xi \leq \frac{1}{6} \ .
\eeq
The solution of eqn.(\ref{varphi_soln}) involves elliptic integrals. Rather 
than present the exact, but cumbersome, expressions, we choose to display 
the profile of $\varphi(z)$ in Fig.1. Since the value of 
$\varphi(z)$ is not of any physical relevance, we have fixed the 
constant of integration such that $\varphi(0) = 0$. 
As is quite apparent, 
the variation of ${\cal V}(\varphi)$ with $\varphi$ is not a rapid one, and 
the bulk of the energy density stored in $\varphi$ is on account of 
the rapidly varying metric. What is also reassuring is that the dependence 
of $\rho(z)$ on the parameter $\xi$ is not extreme.

\begin{figure}[!h]
\vspace*{-13ex}
\epsfig{file=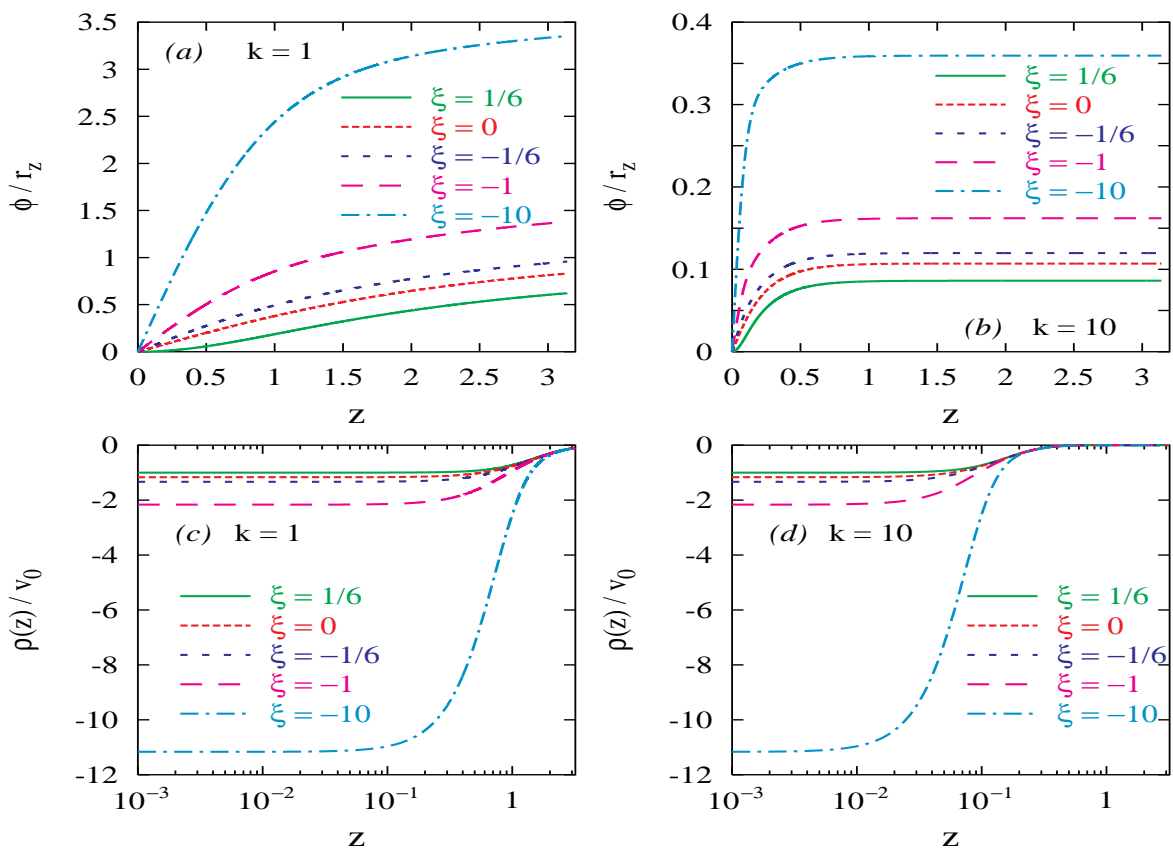,width=17cm,height=17cm}
\vspace*{-21ex}
\caption{\em The profile of the field $\varphi(z)$ (upper panels) 
on the 4-brane at 
$y= \pi$ as well as the corresponding potentials (lower panels) 
$\rho(z)$. The field is defined so that $\varphi(0) = 0$. Left (right) 
panels correspond to $k = 1 \, (10)$.}
  \label{Fig:phi}
\end{figure}
\begin{figure}[!h]
\vspace*{-4ex}
\epsfig{file=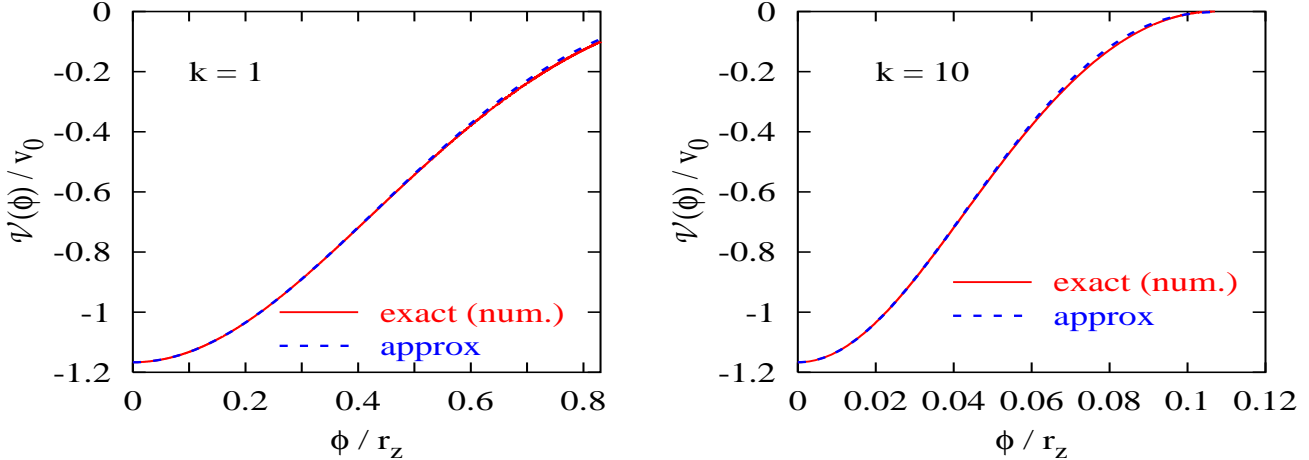,width=17cm,height=7.5cm}
\vspace*{-6ex}

\caption{\em The potential ${\cal V}(\varphi)$ 
on the 4-brane at 
$y= \pi$ as function of $\varphi$ for $\xi = 0$. The solid (red) 
curve is the exact numerical solution while the dashed (blue) curve 
is the approximation of eqn.(\protect\ref{approx_xi0}). 
Left (right) 
panels correspond to $k = 1 \, (10)$.}
  \label{Fig:approx}
\end{figure}

It may be observed from 
eqns.(\ref{def_rho}, \ref{rho_1} \& \ref{varphi_soln})
that although the potential for the scalar field $\cal V$ and the scalar field 
$\phi$ are expressed in terms of the compact coordinate $z$, it is, 
in general, extremely difficult to 
invert the relation and express the scalar potential $\cal V$ 
in terms of $\phi(z)$ through some algebraic equation. This, however, 
can be achieved in certain limits. For example, in the 
$\xi \to - \infty$ limit, we obtain
\[
\beta (\varphi - \varphi_0) = \tanh(kz)
\]
where $\beta^2 = -v_0 \, \xi \, r_z^2 /k^2$ 
and $\varphi_0$ is an integration constant. This, of course, is 
reminiscent of a kink solution. The corresponding  
scalar potential is
\[
{\cal V}(\varphi) = -\alpha v_0 [1 - \beta^{2} (\varphi - \varphi_0)^2]^2 \ .
\]
While this may seem to represent a potential unbounded from
below, note that the solution need not be a runaway one. 
Rather, $\phi = 0$ 
is a deep local minimum of this solution, and the classical 
configuration described above stretches from this minimum to a
point far short of the summit that $\phi$ would need to cross 
to be able to reach the runaway global minimum. It should also 
be borne in mind that such potentials are not uncommon in 
effective field theories in general, and, in particular the low 
energy actions derived from string theory which perhaps is the 
best candidate for ultraviolet completion of such theories that 
we are concerned with. It should also be realised that large $|\xi|$
is not the only scenario wherein a closed form solution can be expected,
but perhaps is the simplest one. 
In the opposite limit, namely  $\xi \to 0$,
eqn.(\ref{varphi_soln}) yields
\beq
\varphi' = A \; \sqrt{{\rm sech}(k \, z)} \ , 
  \qquad A \equiv \sqrt{\frac{v_0 \, r_z^2}{6}} \ .
      \label{xi_eq_0}
\eeq
Integrating this, we have, for large $|k \, z|$, 
\[
exp(-kz) \approx \frac{3 \, k}{ 4 \, v_0 \, r_z^2} (\varphi - \varphi_0)^2 \ , 
\quad
V(\varphi) \approx -\frac{7}{4 \, r_z^2} \, (\varphi -\varphi_0)^2 
\]
where $\varphi_0$ is an integration constnt. A much more interesting 
solution can be obtained by expanding eqn.(\ref{xi_eq_0}) around $z = 0$,
namely
\[
 \varphi'(z) \approx   A  \, \left[1 + \frac{k^2 \, z^2}{4}\right]^{-1/2}
\]
to yield
\beq
\varphi(z) \approx \frac{2 \, A}{k} \, \tan^{-1} \frac{k \, z}{2} 
\ ,
\qquad
{\cal V}(\varphi) \approx \frac{-7 \, v_0}{6} \; 
                          {\rm sech} \left( 2 \, \tan \frac{k \, \phi}{2 \, A}
                                     \right)
   \label{approx_xi0}
\eeq
Note that the above potential is a periodic one! 
In Fig.\ref{Fig:approx}, we compare it with the exact numerical solutions
presented in Fig.\ref{Fig:phi} for the $\xi = 0$ case. The remarkable 
agreement bears testimony to the goodness of the approximation in 
eqn.(\ref{approx_xi0}), which is not surprising since it 
also analytically matches with the approximate solution obtained above for 
large $k \, z$. For non-zero finite values of $\xi$, a good approximate 
solution is admittedly more difficult to obtain, but the above examples 
illustrate that it may  not be impossible to!

While, for the 4-brane at $y = \pi$, the 
$z$-dependent brane energy density can easily be
accounted for in terms of the scalar $\varphi(z)$, a
similar analysis for the brane situated at $y= 0$, 
leads to a somewhat different conclusion.
For this ($y=0$) brane we have 
\beq \dis
\rho = v_0 \; \left[ \frac{7}{6} \, {\rm sech}(k \, z) 
                        + \tilde \xi \; {\rm sech}^4(k \, z) \right] \ ,
\qquad 
\frac{\varphi'^2}{r_z^2} = v_0 \; \left[ \frac{-1}{6} \,
{\rm sech}(k \, z) + \tilde \xi \; {\rm sech}^4(k \, z)
\right] \ .
    \label{phi_phant}
\eeq 
Once again, positivity of $\varphi'^2$ requires
\beq \tilde \xi \geq  \frac{1}{6} \; \cosh^3(k \, \pi)  \ .
    \label{xitil}
\eeq 
and as in the previous case, the large $\tilde \xi$ limit yields the potential for
the scalar field $\cal V(\phi)$ as,
\beq
{\cal V}(\varphi) = \tilde \xi v_0 [1 - \beta^{2} (\varphi - \varphi_0)^2]^2
\eeq   
It may be observed that in this case the potential is not unbounded
from below.  Proceeding similarly, one can find the form of the scalar
potential in small $|\xi|$ regime also.  Several comments are in order
here:
\begin{itemize}
\item If the hierarchy between the Planck scale and the TeV scale is to
  be explained primarily by the warping in th $z$-direction, then
  $\cosh (k \, \pi)$ is large and eqn.(\ref{xitil}) implies a very large
  value for the parameter $\tilde \xi$. Although it might be argued
  that this unnaturalness is just a consequence of the particular
  parametrization of $\varphi(z)$ on this brane, the large difference 
  between $\xi$ and $\tilde \xi$ is indeed disquieting.

\item A (drastic) way out of this would be to exchange the field $\varphi(z)$
  (on this particular brane) for a phantom scalar field (i.e., one whose 
  kinetic term has the opposite sign). This, obviously, would necessitate
  $\tilde \xi \leq 1/6$ rather than eqn.(\ref{xitil}). 
  
The presence of a phantom field in the theory does not necessarily
imply a discernible role for it on the SM 3-brane. However, if we
identify the latter with the one located at $(y = 0, \, z = 0)$---see
Sect.\ref{brane-id}---then this raises the interesting possibility of
obtaining a dark energy candidate with a non-trivial equation of
state.

However, as is well-known, such a scalar field is not admissible in a
fundamental theory. Thus, invoking such a course would necessitate considering 
the present theory as an  effective field theory description of a different 
theory. Though this, admittedly, is somewhat counterintuitive in a theory 
purporting to be valid until $M_{P}$, yet such an eventuality cannot be 
ruled out in principle.

\item A possible alternative to a phantom-like nature for $\varphi(z)$ would 
be to postulate a non-minimal coupling of the same to gravity on the brane,
thereby effecting a change in both of eqns.(\ref{phi_phant}). This, however, 
needs further investigation.

\item Perhaps the simplest way around eqn.(\ref{xitil}) is to appeal to the fact
that if we demand that the warping in the $y$-direction is to account for the 
Planck scale--TeV scale hierarchy, then $c$ is large and $k$ is small
(see Sect.\ref{brane-id}). This, in turn, implies that not too large a value 
for $\tilde \xi$ can still ensure positivity of $\varphi'^2$. 
\end{itemize}

While the alternatives listed above present several possible solutions to the 
problem of a $z$-dependent energy density concentrated on the 4-brane at 
$y = 0$, it should be realized that each will have its own unique set of 
phenomenological consequences (and, in the case of one, require an ultraviolet
completion). We postpone any such discussion to a future occasion and turn
instead to a brief examination of some outstanding issues.

The braneworld model proposed here faces the usual problem of
stability of the moduli $r_z$ and $R_y$, that one encounters even in
the original 5 dimensional Randall-Sundrum model. To stabilize the
single modulus $r_c$ in that model, the most well known mechanism was
formulated by Goldberger and Wise where a bulk scalar field is used to
stabilize the brane separation. Here, one may carry a similar analysis
by incorporating a six dimensional bulk scalar field so that the
back-reaction of the scalar field on the background metric is
negligibly small. By integrating out the scalar field, an effective
potential for the moduli can be generated. The minimization conditions
of the potential would give the stabilized value of the moduli. Since the 
corresponding solutions involve several hypergeometric functions and elliptic
integrals, we desist from presenting them here.

Finally, because of the flat nature of the metric on the 3-branes, the
induced cosmological constant on the TeV-brane, as in the RS case,
vanishes identically.

\section{Seven and higher dimensional spacetime with multiple warping} 

In our task of extending our solutions to even higher dimensions, we 
start with a seven dimensional spacetime wherein three
dimensions are successively warped. In other words, the 
manifold of interest is 
$\left[ \left\{M^{(1, 3)} \times [S^1 (/ Z_2)] \right\}
        \times [S^1 (/ Z_2)] \right] \times [S^1 (/ Z_2)]$.
As in Sec.\ref{sec:6d}, the total bulk-brane action is given by,
\beq
\barr{rcl}
S & = & \dis S_7 + S_6 + S_5 + S_4 \\[1ex]
S_7 & = & \dis \int {d^4 x} \, {d y} \, {d z} \, {d w} \, 
          \sqrt{-g_7} \; \left(R_7 - \Lambda_7 \right) 
\\[2ex]
S_6 & = & \dis \int {d^4 x} \, {d y} \, {d z} \,  {d w} \, 
           \left[ V_1 \, \delta(w ) + V_2 \, \delta(w - \pi) \right]
\\[1.5ex]
    & + & \dis \int {d^4 x} \, {d y} \, {d z} \,  {d w} \, 
           \left[ V_3 \, \delta(z ) + V_4 \, \delta(z - \pi) \right]

\\[1.5ex]
    & + & \dis \int {d^4 x} \, {d y} \, {d z} \,  {d w} \, 
           \left[ V_3 \, \delta(y ) + V_4 \, \delta(y - \pi) \right]
\earr
\eeq
with appropriate actions ($S_5$) for twelve possible  4-branes at 
the edges $(z, w) = (0,0), (0, \pi), (\pi, 0), (\pi, \pi)$,\\
$(z, y) = (0,0), (0, \pi), (\pi, 0), (\pi, \pi)$
and $(y, w) = (0,0), (0, \pi), (\pi, 0), (\pi, \pi)$
and eight 
possible  3-branes at the corners, $(y, z, w) = (0,0, 0), (0, 0, \pi), (0, \pi, 0), 
(0, \pi, \pi), (\pi,0, 0), (\pi, 0, \pi), (\pi, \pi, 0)$ and 
$(\pi, \pi, \pi)$. 
As a natural extension to our previous result we make the following metric ansatz:
\beq
d s^2 =  f^2(w)\, \left[ 
    b^2(z)\left\{a^2(y)\eta_{\mu \nu} d x^\mu d x^\nu 
      + R_y^2 \, d y^2\right\} + r_z^2 \, d z^2 \right] + d w^2\\
\eeq
where $\eta_{\mu \nu} = diag(-1, 1,1,1)$.
Solving  Einstein's equation, one obtains, for 
the metric coefficients $a(y), b(z)$ and $f(w)$,
\beq 
\barr{rcl}
a^2(y) & = & \dis e^{2 \, c \, y} 
\\[2ex]
b^2(z) & = & \dis 
\left\{  
\barr{rcl}
b^2_1(z) & = & \dis 
       \frac{c^2}{k^2 \, R^2_y} \, \cosh^2[ k \, r_z \, (z - z_0)] 
 \\[2ex]
b^2_2(z) & = & \dis 
       \frac{- c^2}{k^2 \, R^2_y} \, \sinh^2[ k \, r_z \, (z - z_0)]
\earr
\right.
\\[4ex]
f^2(w) & = & \dis 
\left\{  
\barr{rcl}
f^2_1(w) & = & \dis 
       \frac{15 \, k^2}{\Lambda_7} \, 
             \cosh^2\left[ \sqrt{\frac{\Lambda_7}{15} }  \, \rw \, (w - w_0) 
                  \right]
 \\[2ex]
f_2(w) & = & \dis 
        - \frac{15 \, k^2}{\Lambda_7} \, 
             \sinh^2\left[ \sqrt{\frac{\Lambda_7}{15} }  \, \rw \, (w - w_0) 
                  \right]
\earr
\right. \ ,
\earr
 \label{soln:7d}
\eeq
where we have assumed that $c^2 > 0$. 

Note that functions of the form $b(z) \sim e^{\alpha z}$,
or $f(w) \sim e^{\beta w}$ are not solutions.
As eqn.(\ref{soln:7d}) suggests, there are four materially 
different solutions in the bulk,
$[ f_i(w), b_j(z) ], \ i, j = 1,2$. 
However, it is easy to see that for three 
of the four combinations, the seven dimensional spacetime is
endowed with two timelike and five spacelike direction. Discarding 
such solutions, we are left with only 
$f(w) = f_1(w)$, $b(z) = b_1(z)$, or in other words, 
\beq 
\barr{rcl}
d s^2 & = & \dis \frac{\cosh^2 (\ell \, w)}{\cosh^2 (\ell \, \pi)} \,
          \left\{ \frac{\cosh^2(k \, z)}{\cosh^2 (k \, \pi)} \,
    \left[ \exp\left(- 2 \, c \, y \right)
                 \, \eta_{\mu \nu} d x^\mu d x^\nu 
      + R_y^2 \, d y^2 \right] + r_z^2 \, d z^2 \right\} + \rw^2 \, d w^2
\\[3ex]
\ell^2 & \equiv & \dis \frac{\Lambda_7 \, \rw^2}{15} 
\\[2ex]
k & \equiv & \dis \frac{\ell \, r_z}{\rw \, \cosh (\ell \, \pi)}
\\[3ex]
c & \equiv & \dis \frac {\ell \, R_y}
                        {\rw \, \cosh(k \, \pi) \, \cosh(\ell \, \pi)}
    \,  =  \, \frac {k \, R_y}
                        {r_z \, \cosh(k \, \pi) } \ .
    \label{soln_1}
\earr
\eeq
As before, the factors of $\cosh (\ell \, \pi)$ and $\cosh (k \, \pi)$
in the metric are included to ensure that the natural scale never 
surpasses unity. 


It may be observed that the 5-brane at $w = \pi$ does not have a flat
metric ($y$- and $z$-dependences).  Now, to obtain substantial
warping in the $w$-direction (from $w = \pi$ to $w = 0$), one would
need $\ell \pi$ to be substantial (same order of magnitude as the
usual Randall-Sundrum case). 
However, this immediately means that
both $k$ and $c$ in eqn.(\ref{soln_1}) are small (for $r_z, R_y \sim
\rw$).  Which, in turn, implies that we cannot have a large warping in
either of $y$- and $z$-directions. 
Of course, if we do not demand a very large warping in $w$ [
$\cosh(\ell \, \pi) \sim {\cal O}(1)$, or, in other words, $\ell \,
\pi \sim {\cal O}(1)$ ], then we can have a large warping in $z$ (or $y$). 

The seven-dimensional (triply-warped) 
theory, then, has a structure very analogous 
to that of the six-dimensional (doubly-warped) one, 
not only in the functional dependence of the metric, but also as far as
the extent of warping is concerned. As can easily be recognised, the 
solution can be almost trivially extended to even higher dimensions.

Note that orbifolding demands that we have to have branes
situated at the edges of the $n$-dimensional hypercube, and possibly
3-branes at the corners.  Now, if one direction (say $z_1$) suffers
from a large warping, then those in the other directions are
necessarily small. This, then, leads to a situation where all the
3-branes at the same $z_1 (=z_1^0)$ coordinate as ours must have a
natural scale relatively close to ours (TeV), although still separated
from us by the small warpings in the ($n-1)$ directions orthogonal to
us. In other words, if we have SM-like fields in each of these
3-branes, the apparent mass-scales (on each brane) would be close to
TeV with some splittings. This leads to a phenomenologically
interesting possibilities which we discuss in the following section.

\section{Some Phenomenological Consequences}
\subsection{Fermion Masses}
We now speculate on some possible phenomenological consequences and
constructs. The hierarchy among the masses of the standard model
fermions has been a subject of interest for a long time. There have
been various efforts in this direction through scenarios like
radiative corrections, different grand unification schemes
etc.~\cite{fermionmass}. In a slightly different context of a
universal extra dimensional model it has been shown\cite{dobrescu}
that the requirement of anomaly cancellation in presence of 
two extra dimensions constrains the number of fermion generations in standard
model to three.
We now explain how our model of multiple warped
geometry can give rise to the observed mass splitting in these standard
model fermions.  As we have seen in Section 2, in a 6-dimensional
doubly warped scenario, the extent of warping in the two directions
are, in general, very different.  In fact, if warping were to explain
the large hierarchy between the Planck scale and the apparent scale
for the electroweak interactions (namely the TeV scale), then the two
warpings necessarily have to be very different in magnitude. In other
words, we have a situation such that there are two branes close to the
Planck scale with two more being at the electroweak scale.  And the
second TeV-like brane could as well have a natural scale slightly
below us as above us. In other words, if we have SM-like fields in
each of these 3-branes, the apparent mass-scales (on each brane) would
be close to TeV with some splitting between them.

Now, imagine the SM fermions being defined by 5-dimensional fields, restricted
to the 4-brane at $z = 0$, which now defines the ``bulk'' for these
fields. If the major warping has occurred in the $z$-direction, then
the natural mass scale of these fields is still ${\cal O}({\rm TeV})$.
The presence of a $y$-dependence in the metric obviously leads to a
non-trivial bulk wavefunction. Furthermore, since this 4-brane also
intersects two other 4-branes at $y = 0$ and $y = \pi$ respectively,
on the resultant 3-branes, the fermion fields are allowed have
brane-kinetic terms in addition to the bulk kinetic term~\cite{opaque}. 
The presence
of such boundary kinetic terms immediately alters the fermion
wavefunction in the bulk (4-brane) as well as on the 3-branes. This,
in turn, changes the overlap of the fermion wavefunction with that of
a scalar located on the 3-brane and thus the effective Yukawa
coupling. Note that the brane kinetic term is the resultant of
interactions of the given fermion field with the other fields on the
brane. Thus, slightly differing interactions on the distant 3-brane
would result in a hierarchy amongst the effective Yuakawa couplings on
our 3-brane and hence the fermion masses.

It is easy to see that this feature is repeated in the case of
higher-dimensional ($d = 4 + n$) constructs (Sections 3, 4). In
addition, certain other features may also appear.  Note that
orbifolding demands that we have to have branes situated at the faces 
and edges
of the $n$-dimensional hypercube, and possibly 3-branes at the
corners.  Once again, if one direction (say $z_1$) suffers from a
large warping, then those in the other directions are necessarily
small. This, then, leads to a situation where all the 3-branes at the
same $z_1 (=z_1^0)$ coordinate as ours must have a natural scale
relatively close to ours (TeV), although still separated from us by
the small warpings in the ($n-1)$ directions orthogonal to us.  Now
consider the different SM fermion fields to be higher-dimensional
ones, but confined to {\em different} $p$-branes, which are all
situated at $z_1 = z_1^0$ and intersect to give our 3-brane.  For each
such fermion, the corresponding $p$-brane defines the bulk. On account
of the slightly different warping on each of these $p$-branes. these
fermions will have differing expressions for the wavefunction in the
respective bulk and thus on the SM 3-brane, thereby resulting in a
hierarchy of Yukawa couplings.  Furthermore, the fermions would be
associated with naturally differing brane kinetic energies which, in
turn, leads to further fine-tuning of the Yukawas.
It should be noted that the above is only a plausibility 
argument in favour of a geometrodynamical origin of fermion 
Yukawa masses in a multi-warped universe. A realistic structure 
needs yet to be constructed.

\subsection{Graviton tower}
A different consequence, not necessarily related to the one discussed
above, pertains to the nature of the Kaluza-Klein towers. Assuming,
for simplicity, that the SM fields are confined to our 3-brane alone,
we are faced with just one relevant field, namely the
graviton. Clearly, we have a mutiple, and intertwined, tower in
place. To divine the exact nature of the tower including the spacings
between the modes and the corresponding eigenfunctions, requires us to
solve the graviton equation of motion. This can be done, in the
weak-field limit, a la Randall-Sundrum on effecting some changes in
variables. Although it is obvious that the exact equations are much
more complicated than in the RS case, several qualitative features are
easy to appreciate:
\begin{itemize}
\item If, as has been argued already, the bulk of the hierarchy problem 
      is addressed by the exponential warping in the $y$-direction, then 
      $k$ is small, and the $z$-dependence of the metric is small. This,
      then, reduces the the situation to essentially a RS $\otimes$ ADD 
      one, with the ADD radius being very small. In other words, 
      the graviton tower, as felt by low-energy experiments, 
      would be almost identical to the RS case. 
\item For the opposite case, viz. $k \, \pi \sim 10$, we again have 
      a similar scenario, with the caveat that the graviton wavefunction 
      (and masses) would be changed somewhat compared to the RS case.
\item Especially with the opening of many (upto 6, if we have string theory 
      in mind) extra dimensions, a more interesting possibility opens itself.
      If we allow for a progression of small ($\lsim {\cal O}(10)$) 
      hierarchies between the moduli, then many intermediate scales become 
      available to us. The multiplicity of towers as well as the intermediate 
      scale may become very relevant in collider phenomenology. 
\end{itemize}

\section{Conclusions} 
In summary, the exact solutions of higher dimensional Einstein's
equation for a mutiply warped spacetime with negative cosmological
constant has been found. It is shown that the hierarchy problem can be
resolved geometrically without invoking any further hierarchy among
the various moduli provided the warping is large in one direction and
small in the other. Thus, in the case of a six dimensional spacetime, one of
the compact dimensions is nearly flat while the other is srongly
warped. The resulting geometry is thus similar to a combination RS and
ADD scheme of compactifications.  We have further shown that such a
situation automatically leads to a spectral splitting of scales around
the Planck and TeV scale thereby providing a clue to the mass
splitting of the standard model fermions. The 4-dimensional brane
tension turns out to be dependent on compact coordinates, indicating
the existence of an effective scalar field distribution along the
branes which is expected to have non-trivial effects on the physics in
the bulk. We further speculate that the excitation of bulk fields like
scalar, gravity , gauge and higher form tensor fields alongwith their
appropriate Kaluza-Klein modes may give rise to interesting
phenomenological signatures in our search for extra dimension in the
forthcoming collider experiments.

\section*{Acknowledgments}
DC acknowledges support from the 
Department of Science and Technology, India under project number
SR/S2/RFHEP-05/2006.


\begin{thebibliography}{99}
\bibitem{arkani} N. Arkani-Hamed, S. Dimopoulos and G. Dvali, 
Phys. Lett. {\bf B429} 263 (1998); 
I. Antoniadis, N. Arkani-Hamed, S. Dimopoulos and G. Dvali, 
Phys. Lett. {\bf B436} 257 (1998).

\bibitem{antoniadis} I. Antoniadis, Phys. Lett. {\bf B246} 377 (1990); 
J.D. Lykken, Phys. Rev. {\bf D54} 3693 (1996); 
R. Sundrum, Phys. Rev. {\bf D59} 085009 (1999); 
K.R. Dienes, E. Dudas and T. Gherghetta, Phys. Lett. {\bf B436} 55 (1998); 
G. Shiu and S.H. Tye, Phys. Rev. {\bf D58} 106007 (1998); 
Z. Kakushadze and S.H. Tye, Nucl. Phys. {\bf B548} 180 (1999).

\bibitem{witten} P. Horava and E. Witten, Nucl. Phys. {\bf B475} 94 (1996); 
{\it ibid} {\bf B460} 506 (1996).

\bibitem{randall}  L. Randall and R. Sundrum, 
Phys. Rev. Lett. {\bf 83} 3370 (1999); 
{\it ibid} {\bf 83} 4690 (1999).

\bibitem{lykken} N. Arkani-Hamed, S. Dimopoulos, G. Dvali and N. Kaloper, 
Phys. Rev. Lett. {\bf 84} 586 (2000); 
J. Lykken and L. Randall, JHEP {\bf 06} 014 (2000);
C. Csaki and Y. Shirman, Phys. Rev. {\bf D61} 024008 (2000); 
A.E. Nelson, Phys. Rev. {\bf D63} 087503 (2001).

\bibitem{cohen} A.G. Cohen and D.B. Kaplan, hep-th/9910132; 
C.P. Burgess, L.E. Ibanez and F. Quevedo, Phys. Lett. {\bf B447} 257 (1999); 
A. Chodos and E. Poppitz, Phys. Lett. {\bf B471} 119 (1999); 
T. Gherghetta and M. Shaposhnikov, Phys. Rev. Lett. {\bf 85} 240 (2000).

\bibitem{kaloper} N. Kaloper, Phys. Rev. {\bf D60} 123506 (1999); 
T. Nihei, Phys. Lett. {\bf B465} 81 (1999); 
H.B. Kim and H.D. Kim, Phys. Rev. {\bf D61} 064003 (2000).


\bibitem{gw} W.D. Goldberger and M. B. Wise, 
   Phys. Rev. Lett. {\bf 83} 4922 (1999); Phys. Lett. {\bf B475} 275 (2000)

\bibitem{csaki}O. DeWolfe, D.Z. Freedman, S.S. Gubser and A. Karch, 
Phys. Rev. {\bf D62} 046008 (2000); 
C. Csaki, M.L. Graesser and G. D. Kribs, Phys. Rev. {\bf D63} 065002 (2001);
C. Csaki, M.L. Graesser, L. Randall and J. Terning, 
Phys. Rev. {\bf D62} 045015 (2000).

\bibitem{ssg}  S. Das, A. Dey and S. SenGupta, 
Class.Quant.Grav.{\bf 23} L67 (2006); 
H. Yoshiguchi {\it et al}, JCAP {\bf 0603} 018 (2006); 
E.E. Boos {\it et al}, hep-th/0511185; 
R. Maartens, Living Rev. Rel. {\bf 7} 7 (2004) and references therein;
D.Maity, S.SenGupta and S. Sur, hep-th/0604195 and hep-th/0609171. 

\bibitem{ferreira} P.C. Ferreira and P.V. Moniz, hep-th/0601070; 
hep-th/0601086; 
G.L. Alberghi {\it et al}, Phys. Rev. {\bf D72} 025005 (2005);
G.L. Alberghi and A. Tronconi, Phys. Rev. {\bf D73} 027702 (2006); 
A.A. Saharian and M.R. Setare, Phys. Lett. {\bf B552} 119 (2003).

\bibitem{bulk} W.D. Goldberger and M.B. Wise, 
Phys. Rev. {\bf D60} 107505 (1999);
S. Kachru, M.B. Schulz and E. Silverstein, Phys. Rev. {\bf D62} 045021 (2000); 
H.A. Chamblin and H.S. Reall, Nucl. Phys. {\bf B562} 133 (1999); 
C. Csaki, hep-ph/0404096; 
R. Neves, TSPU Vestnik {\bf 44N7} 94 (2004); 
E. Dudas and M. Quiros, Nucl. Phys. {\bf B721} 309 (2005). 

\bibitem{cvetic} K. Behrndt and M. Cvetic, Phys. Lett. {\bf B475} 253 (2000).

\bibitem{rshigh} S. Randjbar-Daemi and M.E. Shaposhnikov, 
Phys.Lett.{\bf B491} 329 (2000); 
P. Kanti, R. Madden and K.A. Olive, Phys.Rev.{\bf D64} 044021 (2001); N.Kaloper, JHEP {\bf 0504} 061 (2004);
T.Gherghetta, A.Kehagias, Phys.Rev.Lett {\bf 90} 101601 (2003). 

\bibitem{papantonopoulos} B.Cuadros-Melgar, E.Papantonopoulos, Phys.Rev.{\bf D72} 064008 (2005).


\bibitem{fermionmass} B.S. Balakrishna, Phys.Rev.Lett {\bf 60} 1602 (1988); 
S.M. Barr, Phys. Rev. {\bf D21} 1424(1980);
H. Naoyuki and  Y. Shimizu, hep-ph/0210146.

\bibitem{dobrescu} B.A. Dobrescu and E. Poppitz, Phys.Rev.Lett. {\bf 47} 031801 (2001).
\bibitem{opaque}
  M.~Carena, T.~M.~P.~Tait and C.~E.~M.~Wagner,
  Acta Phys.\ Polon.\ B {\bf 33}, 2355 (2002)
  [arXiv:hep-ph/0207056];
  M.~Carena, E.~Ponton, T.~M.~P.~Tait and C.~E.~M.~Wagner,
  Phys.\ Rev.\ D {\bf 67}, 096006 (2003)
  [arXiv:hep-ph/0212307].

\end{thebibliography}
\end{document}